\documentclass[aps,prd,twocolumn,aps,floatfix,showpacs,tightenlines,showkeys,superscriptaddress,amsmath,amssymb,nofootinbib]{revtex4-1}
\usepackage{amssymb,amsbsy,epsfig,color,graphicx}
\usepackage{color}
\usepackage{[longtable}
\usepackage{array}
\usepackage{dcolumn}   
\usepackage{cellspace}
\usepackage{mathtools}
\usepackage{amstext}
\usepackage{amssymb}
\usepackage{stmaryrd}
\usepackage{stackrel}
\usepackage{graphicx}
\usepackage{esint}
\usepackage[utf8]{inputenc}
\usepackage{blindtext}
\usepackage{float}
\restylefloat{table}
\usepackage{booktabs}
\usepackage{enumitem} 

\usepackage{etoolbox} 
\usepackage{lipsum} 
\usepackage[capitalize]{cleveref}

\usepackage{multirow}
\usepackage[caption=false]{subfig}
\renewcommand\[{\begin{equation}}
\renewcommand\]{\end{equation}}

\newcommand{\ba}{\begin{eqnarray}}
\newcommand{\ea}{\end{eqnarray}}

\makeatletter

\appto{\appendix}{%
	\@ifstar{\def\theequation@prefix{A.}}%
	{}%
}
\makeatother

\begin{document}

	\title{Classical properties of non-local, ghost- and singularity-free gravity}

	\author{Luca Buoninfante}
	\affiliation{Dipartimento di Fisica "E.R. Caianiello", Universit\`a di Salerno, I-84084 Fisciano (SA), Italy}
	\affiliation{INFN - Sezione di Napoli, Gruppo collegato di Salerno, I-84084 Fisciano (SA), Italy}
	\affiliation{Van Swinderen Institute, University of Groningen, 9747 AG, Groningen, The Netherlands}
	\author{Alexey S. Koshelev}
	\affiliation{Departamento de F\'isica and Centro de Matem\'atica e Aplica\c c\~oes, Universidade da Beira Interior, 6200 Covilh\~a, Portugal}
       \affiliation{Theoretische Natuurkunde, Vrije Universiteit Brussel} 
       \affiliation{The International Solvay Institutes, Pleinlaan 2, B-1050, Brussels, Belgium}
	
	\author{Gaetano Lambiase}
	\affiliation{Dipartimento di Fisica "E.R. Caianiello", Universit\`a di Salerno, I-84084 Fisciano (SA), Italy}
	\affiliation{INFN - Sezione di Napoli, Gruppo collegato di Salerno, I-84084 Fisciano (SA), Italy}
	
	\author{Anupam Mazumdar}
	\affiliation{Van Swinderen Institute, University of Groningen, 9747 AG, Groningen, The Netherlands}
	\affiliation{Kapteyn Astronomical Institute, University of Groningen, 9700 AV, Groningen, The Netherlands}
	

\begin{abstract}
In this paper we will show all the linearized curvature tensors in the infinite derivative ghost and singularity free theory of gravity in the static limit.  We have found that in the region of non-locality, in the ultraviolet regime (at short distance from the source), the Ricci tensor and the Ricci scalar are not vanishing, meaning that we do not have a vacuum solution anymore due to the smearing of the source induced by the presence of non-local gravitational interactions. It also follows that, unlike in Einstein's gravity, the Riemann tensor is not traceless and it does not coincide with the Weyl tensor. Secondly, these curvatures are regularized at short distances such that they are singularity-free, in particular the same happens for the Kretschmann invariant. Unlike the others, the Weyl tensor vanishes at short distances, implying that the spacetime metric becomes conformally flat in the region of non-locality, in the ultraviolet. As a consequence, the non-local region can be approximated by a conformally flat manifold with non-negative constant curvatures. We briefly discuss the solution in the non-linear regime, and argue that $1/r$ metric potential cannot be the solution where non-locality is important in the ultraviolet regime.
\end{abstract}

\maketitle


\section{Introduction}

It is quite surprising that even after more than 100 years from its formulation, the theory of Einstein's general relativity (GR) is still being tested to a very high precision, in spite of the fact that there are still open problems and challenges that make the theory incomplete and inquisitive. An experimental fact is that the GR works extremely well in the infrared (IR) regime (at large distances and at late times)~\cite{-C.-M.}. The recent detection of gravitational waves \cite{-B.-P.} has stamped GR's authority as an excellent theory of gravity at large distances. Albeit,  the theory breaks down in the ultraviolet (UV) regime, or in other words at short distances and small time scales, where both classical and quantum singularities appear. In fact, from a classical point of view as soon as we approach very short distances, we encounter blackhole type ($1/r$-singularity, where $r$ is the distance from the source), and cosmological singularities, while from a quantum point of view GR is not perturbatively renormalizable, so that it loses its predictability in the UV regime.

In fact, Newton's $1/r$-fall of the gravitational potential has been tested only up to a distance of $5.6\times 10^{-5}$ meters,  in energy scale it corresponds to $0.004$eV, in torsion balance experiments \cite{-D.-J.}. It means that there is huge gulf, i.e. roughly $30$ orders of magnitude between $0.004$eV and the Planck scale $M_p\sim 10^{19}$GeV~\footnote{We are working in Natural Units: $\hbar=1=c$.}, where the nature of gravitational interaction has not been established completely.

In the past years there have been many attempts to cure GR in the UV, in particular the most straightforward way is to consider higher derivative actions by introducing the quadratic terms in the curvature like $\mathcal{R}^2,$ $\mathcal{\mathcal{R}_{\mu \nu}}\mathcal{\mathcal{R}^{\mu \nu}}$, $\mathcal{\mathcal{R}_{\mu \nu \rho \sigma}}\mathcal{\mathcal{R}^{\mu \nu \rho \sigma}}$, after all these terms are expected to arise at one-loop in quantum corrections. In 1977, it was shown by Stelle that quadratic curvature gravity in $4$ spacetime dimensions is power-counting renormalizable but, unfortunately, it turns out to be non-unitary because of the presence of a spin-$2$ ghost as a propagating degree of freedom arising from the Weyl term \cite{-K.-S.}. 

However, it has been recently shown that the most generic parity-invariant gravitational action, quadratic in curvature, must contain infinite covariant 
derivatives~\cite{Biswas:2005qr,Biswas:2011ar,Biswas:2016etb}, which can be made ghost free when its propagator is analyzed around a given background~\cite{Biswas:2005qr,Biswas:2011ar,Biswas:2016etb}~\footnote{Such an action has been postulated before in Refs.~\cite{-Yu.-V.,Tomboulis,Tseytlin:1995uq,Siegel:2003vt}}: 
\begin{equation}
\begin{array}{rl}
S= & \displaystyle\frac{1}{16\pi G} \displaystyle\int d^{4}x\sqrt{-g}\left[{ \mathcal{R}}+\alpha\left(\mathcal{R}\mathcal{F}_{1}(\Box_{s})\mathcal{R}\right.\right.\\
& \displaystyle \left.\left.+\mathcal{R}_{\mu\nu}\mathcal{F}_{2}(\Box_{s})\mathcal{R}^{\mu\nu}+\mathcal{R}_{\mu\nu\rho\sigma}\mathcal{F}_{3}(\Box_{s})\mathcal{R}^{\mu\nu\rho\sigma}\right)\right], \label{eq:1}
\end{array}
\end{equation}
where $G=1/M_p^2$ is Newton's gravitational constant, $\alpha \sim (1/M_s)^2$ is a dimensionful coupling, $\Box_{s}\equiv\Box/M_{s}^{2}$, with $M_s$ being the scale of non-locality at which new gravitational interaction should manifest, and the d'Alembertian operator is given by $\Box\equiv g^{\mu\nu}\nabla_{\mu}\nabla_{\nu}$, where $\mu,~\nu=0,1,2,3$, and we work with the mostly positive convention for the metric signature, $(-,+,+,+)$. The three form factors $\mathcal{F}_i$ (reminiscence to any massless theory, such as pion-form factors) contain all the information about the non-local nature of gravity. They are analytic functions of $\Box_s$, and can be generally expressed in terms of Taylor series; $\mathcal{F}_{i}(\Box_{s})=\stackrel[n=0]{\infty}{\sum}f_{i,n}\left(\Box_{s}\right)^{n}$. The higher derivative expansion in terms of $\Box_s$ also resembles capturing the essence of non-locality in string theory to all orders in $\alpha'$ corrections~\cite{polchinski}, in the context of close string field theory~\cite{Witten:1985cc}, for a review see~\cite{Siegel:1988yz}. Note that the non-locality appears only at the level of interaction, and not at the level of a free field propagation without a source. 

The coefficients $f_{i,n}$ are not arbitrary and depend on the specifics of the background geometry in order to avoid perturbative ghosts~\cite{Biswas:2005qr,Biswas:2011ar,Biswas:2016etb}. Around constant curvature backgrounds they can be uniquely fixed (up to a specific choice of an {\it entire function}, which we will explain below) by requiring that in the IR, we consistently recover the predictions of GR, or the predictions of a scalar tensor theory by accommodating an extra scalar pole in the scalar propagating degree of freedom~\cite{Biswas:2013kla,Buoninfante}. Note that the general covariance has been always maintained through out from IR-to-UV.
Just to give an example, in the Minkowski spacetime if we demand that the only physical propagating degree of freedom is the massless transverse-traceless spin-$2$ graviton, then the three form factors in Eq. \eqref{eq:1} are chosen to satisfy: $2{\cal F}_1(\Box_s)+ {\cal F}_2(\Box_s) + 2{\cal F}_3(\Box_s)=0$~\cite{Biswas:2011ar}. Similar constraints on ${\cal F}_i$ can be derived for de Sitter and anti-deSitter backgrounds by following Ref.~\cite{Biswas:2016etb}.

The aim of the current paper is to reveal some of the curvature properties of the above action Eq.~(\ref{eq:1}) for a particular form of the form factors ${\cal F}_i$'s first discussed by Biswas, Gerwick, Koivisto and Mazumdar (BGKM), which would lead to amelioration of blackhole and cosmological singularities in the linearized regime. We will first discuss various properties of the curvatures such as the Ricci scalar/tensor, the Riemann, and the Weyl tensor within the linearized limit and briefly discuss the consequences for a full non-linear theory, and its solution near the UV regime.
 
\section{BGKM gravity}

By considering the metric perturbation around the flat spacetime~\footnote{In this paper we will concentrate on asymptotically Minkowski background, and perturbations around it.}:
\begin{equation}
g_{\mu \nu}(x)=\eta_{\mu \nu}+h_{\mu \nu}(x) \label{eq:2},
\end{equation}
the action in Eq. \eqref{eq:1} can be expanded up to quadratic order in $h_{\mu \nu}$ \cite{Biswas:2011ar}. The graviton propagator in this theory is given by (suppressing the spacetime indices)~\cite{Biswas:2013kla,Conroy:2015nva,Buoninfante}
\begin{equation}
\Pi(-k^2)=\frac{1}{a(-k^2)}\left(\frac{\mathcal{P}^{2}}{k^{2}}-\frac{\mathcal{P}^{0}}{2k^{2}}\right),\label{eq:4}
\end{equation}
where $\mathcal{P}^{2},$ $\mathcal{P}^{0}$ are the so called spin-projector operators,  note that $\Pi_{\scriptscriptstyle GR}=\mathcal{P}^{2}/k^{2}-\mathcal{P}^{0}/2k^{2}$ is the massless GR graviton propagator, and  the analytic function $a(-k^2)$ is such that it does not contain any pole in the finite complex plane, and in the limit $k\rightarrow 0,~a(-k^2)\rightarrow 1$. Indeed, these properties can be satisfied if $a(-k^2)$ is given by the exponential of an {\it entire~function}, which does not contain any pole. One simplest choice is made by BGKM to be~\footnote{Other choices of $a(-k^2)$ exist, see in particular~\cite{Tomboulis,Tomboulis:2015,Modesto}. In particular, we will be concentrating on exponential of an {\it entire~function} for the rest of this paper.}:
\begin{equation}\label{eg}
a(-k^2)=e^{k^2/M_s^{2}},
\end{equation}
which gives rise to 
\begin{equation}
\alpha {\cal F}_{1}(\Box_{s})=-\frac{\alpha}{2}{\cal F}_{2}(\Box_{s})=\frac{a(\Box_{s})-1}{\Box},\,\,\,\,a(\Box_{s})=e^{-\Box/M_s^{2}}\,,
\end{equation}
which assumes ${\cal F}_3=0$ from the constraint equation: $2{\cal F}_1+{\cal F}_2+2{\cal F}_3=0$, without loss of generality, in the sense that the choice does not affect the graviton propagation at the order $\mathcal{O}(h_{\mu\nu}^2)$. For other choices of {\it entire function}, see Ref.~\cite{Edholm:2016hbt}, where the UV and IR properties remain intact, in the IR one recovers the predictions of Einstein's GR. This is the case of a decoupling limit, when $M_s \rightarrow \infty$. While in the UV, for $k\rightarrow \infty$, the propagator is exponentially suppressed, indicating that such a theory will behave better in the UV both at a classical and at a quantum 
level~\cite{-Yu.-V.,Tomboulis,Modesto,Talaganis:2014ida}. The power counting arguments show that the theory becomes super-renormalizable. 

By working in the weak-field regime in the presence of a static point-source of mass $m$ with Dirac-delta distribution, i.e. $m\delta^{(3)}(r)$, the linearized metric reads in isotropic coordinates:
\begin{equation}
ds^{2}=-\left(1+2\Phi\right)dt^{2}+\left(1-2\Phi\right)\left[dr^2+r^2d\Omega^2\right] \label{eq:5},
\end{equation}
where $d\Omega^2=d\theta^2+{\rm sin}^2\theta d\varphi^2$, and the metric potential is derived in Ref.~\cite{Biswas:2005qr,Biswas:2011ar}
\begin{equation}
\Phi(r)= -\frac{Gm}{r}{\rm Erf}\left(\frac{M_sr}{2}\right)\label{eq:6}.
\end{equation}
In the regime $M_sr > 2$, we recover the IR Newtonian result: $-Gm/r$ as expected, while in the UV regime of non-locality $M_sr<2$ the potential is non-singular and tends to be a finite negative constant value $\Phi(r)\rightarrow -A$, where 
\begin{equation}
A\equiv \frac{GmM_s}{\sqrt{\pi}}< 1. \label{const.pot.}
\end{equation}
By taking the derivative of the potential we can obtain the gravitational acceleration, or in other words the force, \cite{Buoninfante}:
\begin{equation}
F(r)= -\frac{Gm}{r^2}\left[{\rm Erf}\left(\frac{M_sr}{2}\right)-\frac{e^{-\frac{M_s^2r^2}{4}}M_s r}{\sqrt{\pi}}\right]\label{eq:7},
\end{equation}
which recovers the Newton's law $-Gm/r^2$ at large distances $M_sr > 2$, while in the UV regime, i.e. non-local regime $M_sr<2$, it goes linearly to zero, 
$$F(r)\rightarrow-\frac{GmM_s^3}{6\sqrt{\pi}}r.$$ The fact that force vanishes at the origin shows the classical aspects of asymptotic freedom, and also signifies that BGKM gravity with Eq.~(\ref{eg}) has a mass-gap given by $M_s$, first shown in Ref.~\cite{Frolov}~\footnote{Such a behavior of the gravitational potential can be probed experimentally by studying the quantum spread of the matter wavefunction~\cite{Buoninfante:2017kgj}, and studying the equivalence of Bell's inequality~\cite{Bose:2017nin}.}. 

In Einstein's GR the metric potential becomes large, i.e. $2|\Phi(r)|=2Gm/r>1$, for  $r<r_{sch}\equiv 2Gm$. Instead, in the case of BGKM the metric potential, Eqs. (\ref{eq:5},\ref{eq:6}), remains weak through out $r< 2/M_s$ as long as  $mM_s<M_p^2$. The last inequality is crucial: it suggests that the size of the non-local region is always larger than the standard Schwarzschild radius $r_{sch}=2Gm=2m/M_p^2$, i.e. $2/M_s> 2m/M_p^2=r_{sch}$. This is illustrated in Fig. 1 comparing the two cases at hand.

The time dependent solution of the action Eq.~(\ref{eq:1}) yields a non-singular bouncing cosmology in a vacuum, a solitonic solution which is absent in Einstein's 
GR~\cite{Biswas:2011ar}. In fact, with conformal matter it is sufficient to keep only infinite derivatives in the Ricci scalar from the full action for a homogeneous cosmological solutions, which gives rise to a {\it non-singular bouncing cosmology} in the ultraviolet, for time $t < 1/M_s$~\cite{Biswas:2005qr,Koivisto,Koshelev:2012qn}~\footnote{Because ${\cal F}_i$s
follow a constraint equation, where ${\cal F}_2 $ can be switched-off without loss of generality. While in a homogeneous background the Weyl term can be made vanishing at the background level.}. The avoidance of cosmological singularity has also been studied in the context of focusing theorems established by  Penrose-Hawking~\cite{Penrose:1964wq,Hawking:1973uf}, using the Raychoudhuri's equation~\cite{Raychaudhuri:1953yv}. The dynamical formation of blackholes has also been studied in this context, and it has been found that the formation of singularity is avoided for thick shell scatterings, see Refs.~\cite{Frolov:2015bia,Frolov:2015usa} for details.
\begin{figure}[t]
	\includegraphics[scale=0.275]{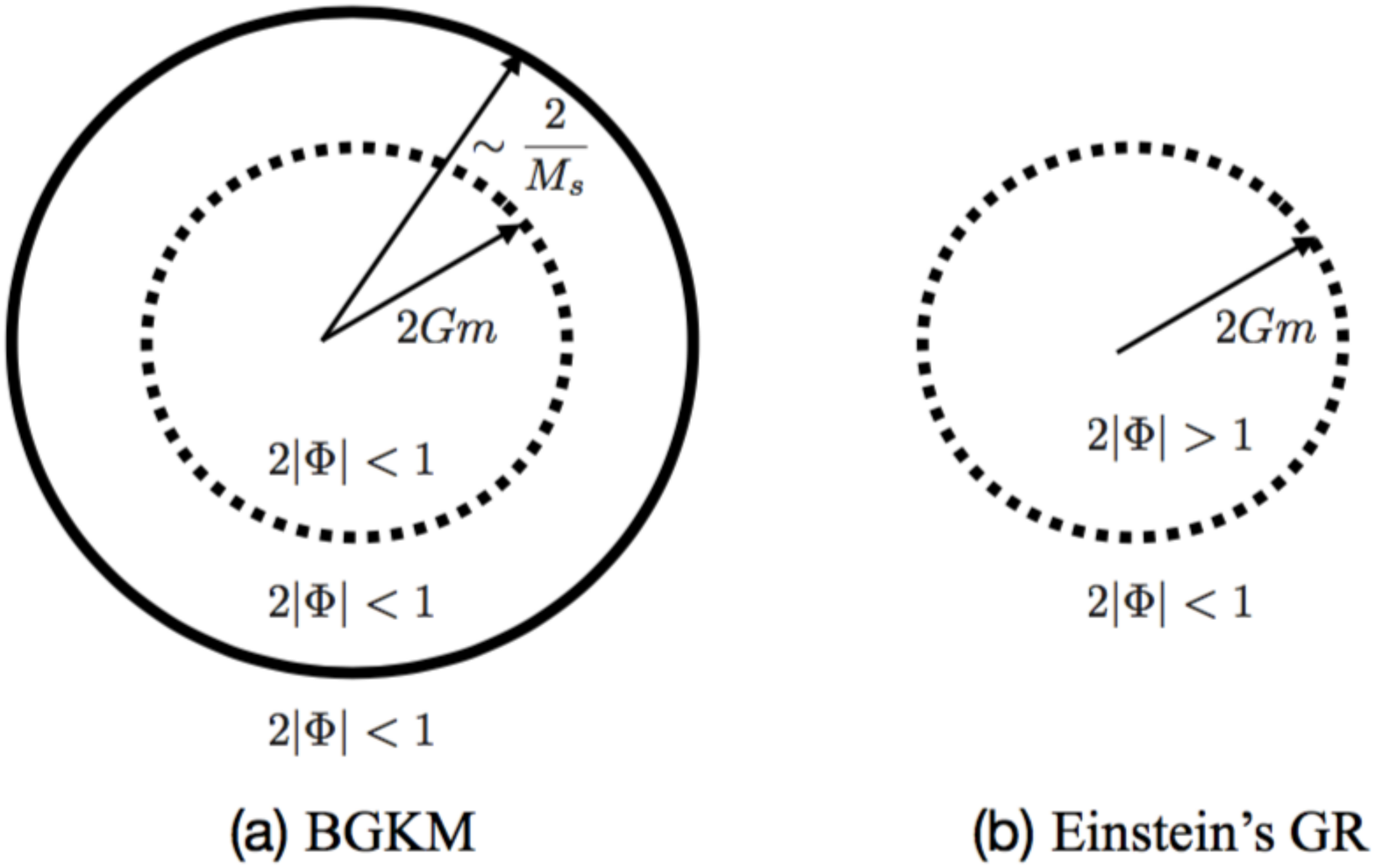}
	\protect\caption{In this figure we have illustrated how the gravitational potential behaves in the cases of (a) BGKM gravity and (b) Einstein's GR. Unlike GR where the metric potential becomes greater than one for distances smaller than the Schwarzschild radius, i.e. $2|\Phi(r)|>1$ if $r<r_{sch},$ in the case of BGKM gravity the metric potential is always smaller than one in the linear regime, i.e. $mM_s<M_p^2 \Longrightarrow 2/M_s>r_{sch}= 2Gm$.}\label{fig1}
\end{figure}

\section{Curvature tensors}

In this section, we wish to compute all the curvature tensors for the metric in Eqs.~(\ref{eq:5},\ref{eq:6}) and make a comparison with the corresponding ones in the Einstein's GR. Although we are working in the linearized regime, by looking at the structure of the curvatures we will be able to say some aspects of the non-local nature of gravity in the non-linear regime too.

We will calculate the components of the Riemann tensor:
\begin{equation}
\begin{array}{rl}
\mathcal{R}_{\mu \nu \rho \sigma}= & \displaystyle \frac{1}{2}\left(\partial_{\nu}\partial_{\rho}g_{\mu \sigma}+\partial_{\mu}\partial_{\sigma}g_{\nu \rho}-\partial_{\sigma}\partial_{\mu}g_{\nu \rho}-\partial_{\mu}\partial_{\rho}g_{\nu \sigma}\right)\\
& \displaystyle \,\,\,\,\,+g_{\alpha\beta}\left(\Gamma^{\alpha}_{\nu\rho}\Gamma^{\beta}_{\mu\sigma}-\Gamma^{\alpha}_{\sigma\nu}\Gamma^{\beta}_{\mu\rho}\right)
\label{eq:8}
\end{array}
\end{equation}
in terms of the  metric in Eq. \eqref{eq:5}, and to be consistent with the linear approximation we will be considering only the pieces linear in the perturbation $h_{\mu\nu}$, or in other words linear in Newton's gravitational constant, as $h_{\mu\nu}\sim G$. In the same way, we will calculate the Ricci tensor $\mathcal{R}_{\mu\nu}=(\eta^{\mu\rho}-h^{\mu\rho})\mathcal{R}_{\mu \nu \rho \sigma}$, the Ricci scalar $\mathcal{R}=(\eta^{\mu\rho}-h^{\mu\rho})\mathcal{R}_{\mu\rho}$, and the Weyl tensor
\begin{equation}
\begin{array}{rl}
\mathcal{C}_{\mu \nu \rho \sigma}= & \displaystyle \mathcal{R}_{\mu \nu \rho \sigma}+\frac{\mathcal{R}}{6}\left(g_{\mu\rho}g_{\nu\sigma}-g_{\mu\sigma}g_{\nu\rho}\right) \\
& \displaystyle -\frac{1}{2}\left(g_{\mu\rho}\mathcal{R}_{\nu\sigma}-g_{\mu\sigma}\mathcal{R}_{\nu\rho}-g_{\nu\rho}\mathcal{R}_{\mu\sigma}+g_{\nu\sigma}\mathcal{R}_{\mu\rho}\right)\;.
\label{eq:9}
\end{array}
\end{equation}
We will also calculate the curvatures squared like $\mathcal{R}^2,$ $\mathcal{R}_{\mu \nu}\mathcal{R}^{\mu \nu}$, $\mathcal{K}=\mathcal{R}_{\mu \nu \rho \sigma}\mathcal{R}^{\mu \nu \rho \sigma}$ and $\mathcal{C}^2=\mathcal{C}_{\mu \nu \rho \sigma}\mathcal{C}^{\mu \nu \rho \sigma}$, where $\mathcal{K}$ is the so called Kretschmann invariant tensor.

Let us briefly recall all the curvature tensors in the case of Einstein's GR for the metric in Eq.~(\ref{eq:5}) with $\Phi(r) = -Gm/r$, and afterwards we will consider the BGKM case. The non-vanishing components of the linearized Riemann tensor in Einstein's GR, or in other words, in the case of a linearized Schwarzschild metric in the isotropic coordinates are given by:
\begin{equation}
\begin{array}{rl}
\displaystyle \mathcal{R}^{\scriptscriptstyle (GR)}_{0101}=-\frac{2Gm}{r^3},\,\,\,\,\,\,\,\,\,\,\,\,\,\,\,\,\,\,\,\, \mathcal{R}^{\scriptscriptstyle (GR)}_{0202}=-\mathcal{R}^{\scriptscriptstyle (GR)}_{1212}=\frac{2Gm}{r},\,\,\,\,\,\,\,& \\
\displaystyle \mathcal{R}^{\scriptscriptstyle (GR)}_{0303}= -\mathcal{R}^{\scriptscriptstyle (GR)}_{1313}=\frac{Gm}{r}{\rm sin}^2\theta,\,\,\,\,\,\,\mathcal{R}^{\scriptscriptstyle (GR)}_{2323}=2Gmr{\rm sin}^2\theta, &
\label{eq:10}
\end{array}
\end{equation}
some of which are clearly singular at the origin. The linearized Ricci tensor and Ricci scalar are vanishing as expected, due to the fact that the Schwarzschild metric is a vacuum solution of Einstein's field equations with a boundary condition at $r=0$ with a non-vanishing source $m$, yields:
\begin{equation}
\mathcal{R}^{\scriptscriptstyle (GR)}_{\mu\nu}=0,\,\,\,\,\,\,\,\mathcal{R}^{\scriptscriptstyle (GR)}=0\,,
\label{eq:11}
\end{equation}
for both linear and non-linear regimes.
The non-vanishing components of the linearized Weyl tensor are given by
\begin{equation}
\begin{array}{rl}
\displaystyle \mathcal{C}^{\scriptscriptstyle (GR)}_{0101}=-\frac{2Gm}{r^3},\,\,\,\,\,\,\,\,\,\,\,\,\,\,\,\,\,\,\,\, \mathcal{C}^{\scriptscriptstyle (GR)}_{0202}=-\mathcal{C}^{\scriptscriptstyle (GR)}_{1212}=\frac{2Gm}{r},\,\,\,\,\,\,\,& \\
\displaystyle \mathcal{C}^{\scriptscriptstyle (GR)}_{0303}= -\mathcal{C}^{\scriptscriptstyle (GR)}_{1313}=\frac{Gm}{r}{\rm sin}^2\theta,\,\,\,\,\,\,\mathcal{C}^{\scriptscriptstyle (GR)}_{2323}=2Gmr{\rm sin}^2\theta, &
\label{eq:12}
\end{array}
\end{equation}
which coincide with the components of the Riemann tensor as expected, since from Einstein's equations in the vacuum it follows that the Riemann tensor is traceless. Moreover, the Ricci squared and the Ricci scalar squared are also vanishing, while the Kretschmann tensor and the Weyl squared get contributions at the order $\mathcal{O}(h_{\mu\nu}^2)$, and they coincide consistently with Eqs. \eqref{eq:10} and \eqref{eq:12}:
\begin{equation}
\mathcal{K}^{\scriptscriptstyle (GR)}=\frac{48G^2m^2}{r^6},\,\,\,\,\,\,\,\,\,\mathcal{C}^{\scriptscriptstyle (GR)2}=\frac{48G^2m^2}{r^6}.
\label{eq:13}
\end{equation}
Note that Eqs. \eqref{eq:10}-\eqref{eq:13} hold true in the linearized regime, when $r>r_{sch}$. However, it is also well known that the full non-linear Schwarzschild metric solution suffers from the singularity at $r=0$, and such a feature is clearly reflected into the structures of the linearized curvatures in Eqs. \eqref{eq:10}-\eqref{eq:13}. In particular, from Eq. \eqref{eq:13} it is very clear that at $r=0$, the Kretschmann invariant blows up, signaling that $r=0$ is a singularity in the Schwarzschild metric in any coordinate system.

\section{Non-vacuum solution in BGKM gravity}

Let us now calculate all the curvature tensors in the case of BGKM. The non-vanishing components of the linearized Riemann tensor are given by:
\begin{equation}
\!\!\!\!\!\!\!\!\!\begin{array}{rl}
\mathcal{R}_{0101}= & \displaystyle \frac{e^{-\frac{M_s^2r^2}{4}}Gm}{2\sqrt{\pi}r^3}\left[4M_sr+M_s^3r^3\right.\\
 & \displaystyle \,\,\,\,\,\,\,\,\,-\left.4\sqrt{\pi}e^{\frac{M_s^2r^2}{4}}{\rm Erf}\left(\frac{M_sr}{2}\right)\right],
\label{eq:14}
\end{array}
\end{equation}
\begin{equation}
\begin{array}{rl}
\mathcal{R}_{0303}= & \displaystyle \mathcal{R}_{0202}{\rm sin}^2\theta\\
= & \displaystyle \frac{Gm{\rm sin}^2\theta}{r}\left[{\rm Erf}\left(\frac{M_sr}{2}\right)-\frac{e^{-\frac{M_s^2r^2}{4}}M_sr}{\sqrt{\pi}}\right],\\
\mathcal{R}_{1313}= & \displaystyle \mathcal{R}_{1212}{\rm sin}^2\theta\\
= & \displaystyle \frac{e^{-\frac{M_s^2r^2}{4}}Gm{\rm sin}^2\theta}{2\sqrt{\pi}r}\left[4M_sr+M_s^3r^3\right.\\
& \displaystyle \,\,\,\,\,\,\,\,\,-\left.4\sqrt{\pi}e^{\frac{M_s^2r^2}{4}}{\rm Erf}\left(\frac{M_sr}{2}\right)\right],\\
\mathcal{R}_{2323}= & \displaystyle 2Gmr{\rm sin}^2\theta\left[{\rm Erf}\left(\frac{M_sr}{2}\right)-\frac{e^{-\frac{M_s^2r^2}{4}}M_sr}{\sqrt{\pi}}\right].
\label{eq:15}
\end{array}
\end{equation}
First of all, note that in the regime $M_sr > 2$ the components in Eqs. \eqref{eq:14}-\eqref{eq:15} of the Riemann tensor recover GR results in Eq. \eqref{eq:10}. In particular, note that in the non-local regime $M_sr<2$, Riemann's components are non-singular, most of them vanish and only the components in Eq. \eqref{eq:14} tend to a finite constant value:
\begin{equation}
\mathcal{R}_{0101}\sim \frac{GmM_s^3}{6\sqrt{\pi}}.
\label{eq:16}
\end{equation}
In Fig. 2.a. the component $\mathcal{R}_{0101}$ has been plotted for both BGKM and Einstein's GR, and it is very clear how non-locality, i.e. $r< 2/M_s$,
regularizes the singularity at the origin.  For the purpose of illustration, we have taken $M_p=1,~m=5,~M_s=10^{-3/2}$, which satisfy the linearity of the metric potential,
Eq.~(\ref{eq:5}). The non-vanishing components of the linearized Ricci tensors and Ricci scalar are given by:
\begin{equation}
\begin{array}{rl}
\mathcal{R}_{00}=\mathcal{R}_{11}= & \displaystyle \frac{e^{-\frac{M_s^2r^2}{4}}GmM_s^3}{2\sqrt{\pi}},\\
\mathcal{R}_{22}=& \displaystyle \frac{e^{-\frac{M_s^2r^2}{4}}GmM_s^3r^2}{2\sqrt{\pi}},\\
\mathcal{R}_{33}=& \displaystyle \frac{e^{-\frac{M_s^2r^2}{4}}GmM_s^3r^2{\rm sin}^2\theta}{2\sqrt{\pi}},
\label{eq:17}
\end{array}
\end{equation}
and
\begin{equation}
\mathcal{R}= \frac{e^{-\frac{M_s^2r^2}{4}}GmM_s^3}{\sqrt{\pi}}.
\label{eq:18}
\end{equation}
In the local regime, or in the IR, where $M_sr\geq2$, the Ricci tensor and the Ricci scalar go to zero consistently with Eq. \eqref{eq:11}, while in the non-local region $M_sr<2$, one obtains:
\begin{equation}
\mathcal{R}_{00}=\mathcal{R}_{11}\sim \frac{GmM_s^3}{2\sqrt{\pi}},\,\,\,\,\,\,\,\,\,\mathcal{R}_{22}=\mathcal{R}_{33}\sim0
\label{eq:19}
\end{equation}
and
\begin{equation}
\mathcal{R}\sim\frac{GmM_s^3}{\sqrt{\pi}}.
\label{eq:20}
\end{equation}
\begin{figure}[b]
	\centering
	\subfloat[Subfigure 1 list of figures text][Riemann component $\mathcal{R}_{0101}$]{
		\includegraphics[scale=0.415]{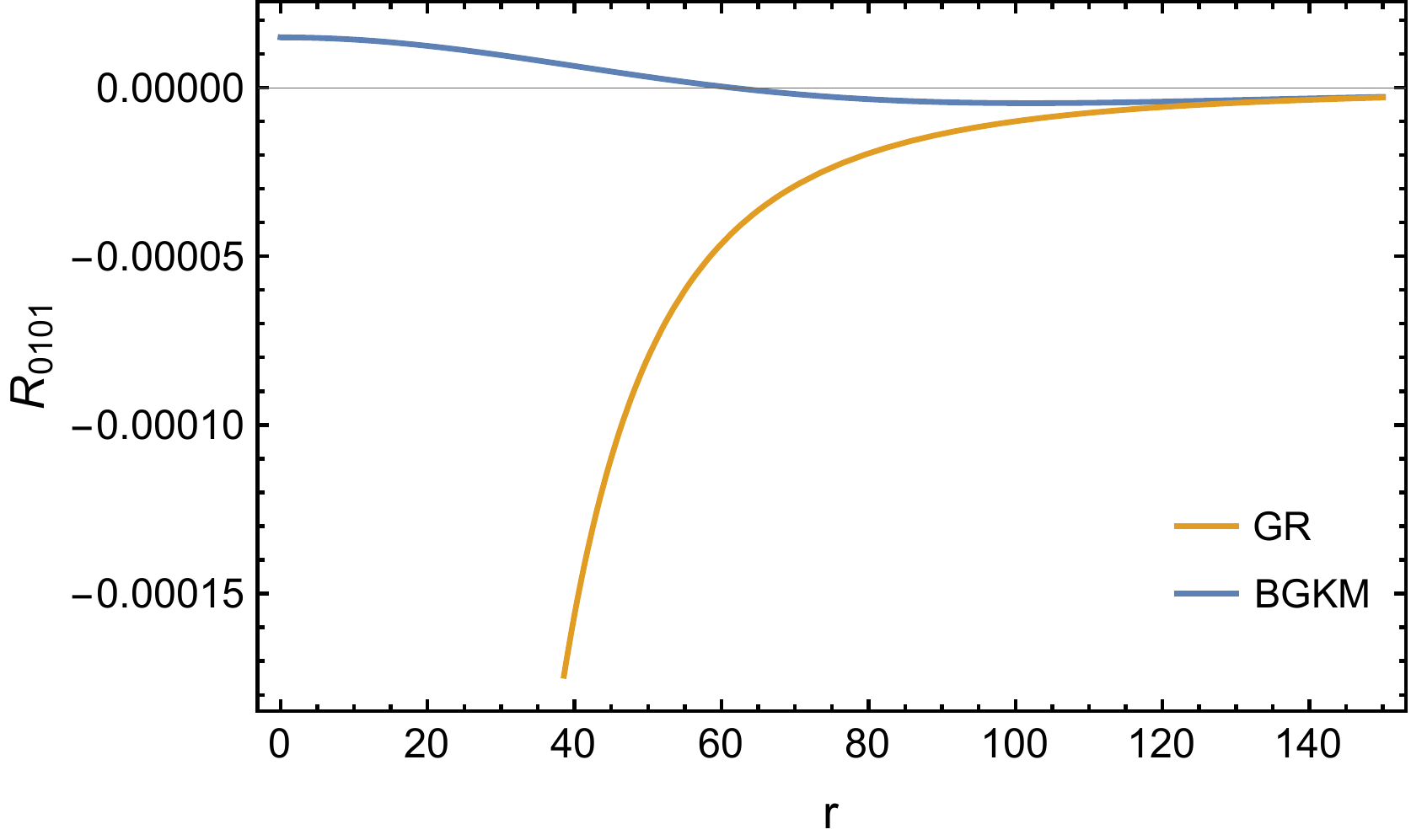}
		\label{fig.2.a}}
	\qquad
	\subfloat[Subfigure 2 list of figures text][Weyl component $\mathcal{C}_{0101}$]{
		\includegraphics[scale=0.42]{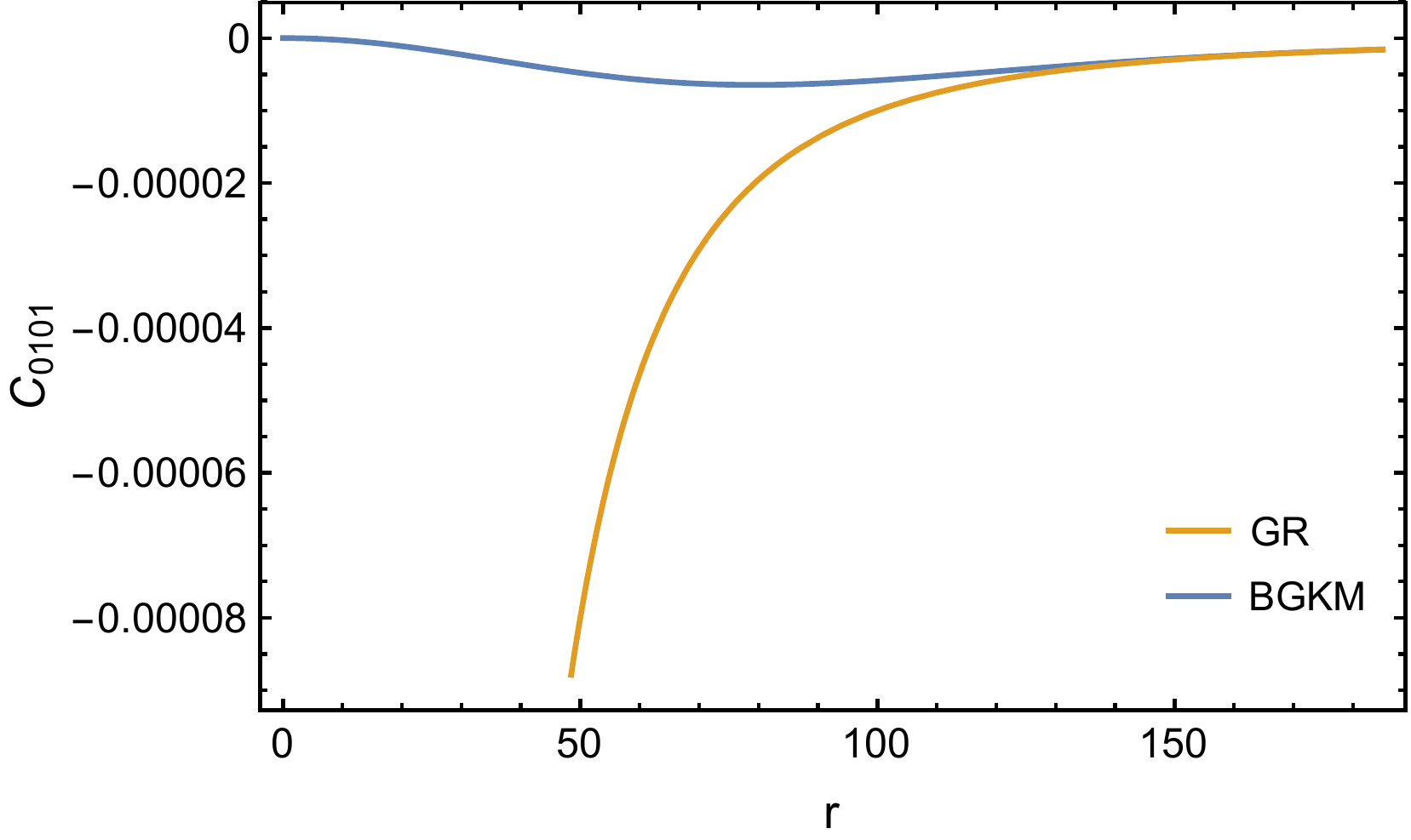}
		\label{fig.2.b}}
	\qquad
	\subfloat[Subfigure 3 list of figures text][Kretschmann invariant $\mathcal{K}$]{
		\includegraphics[scale=0.4205]{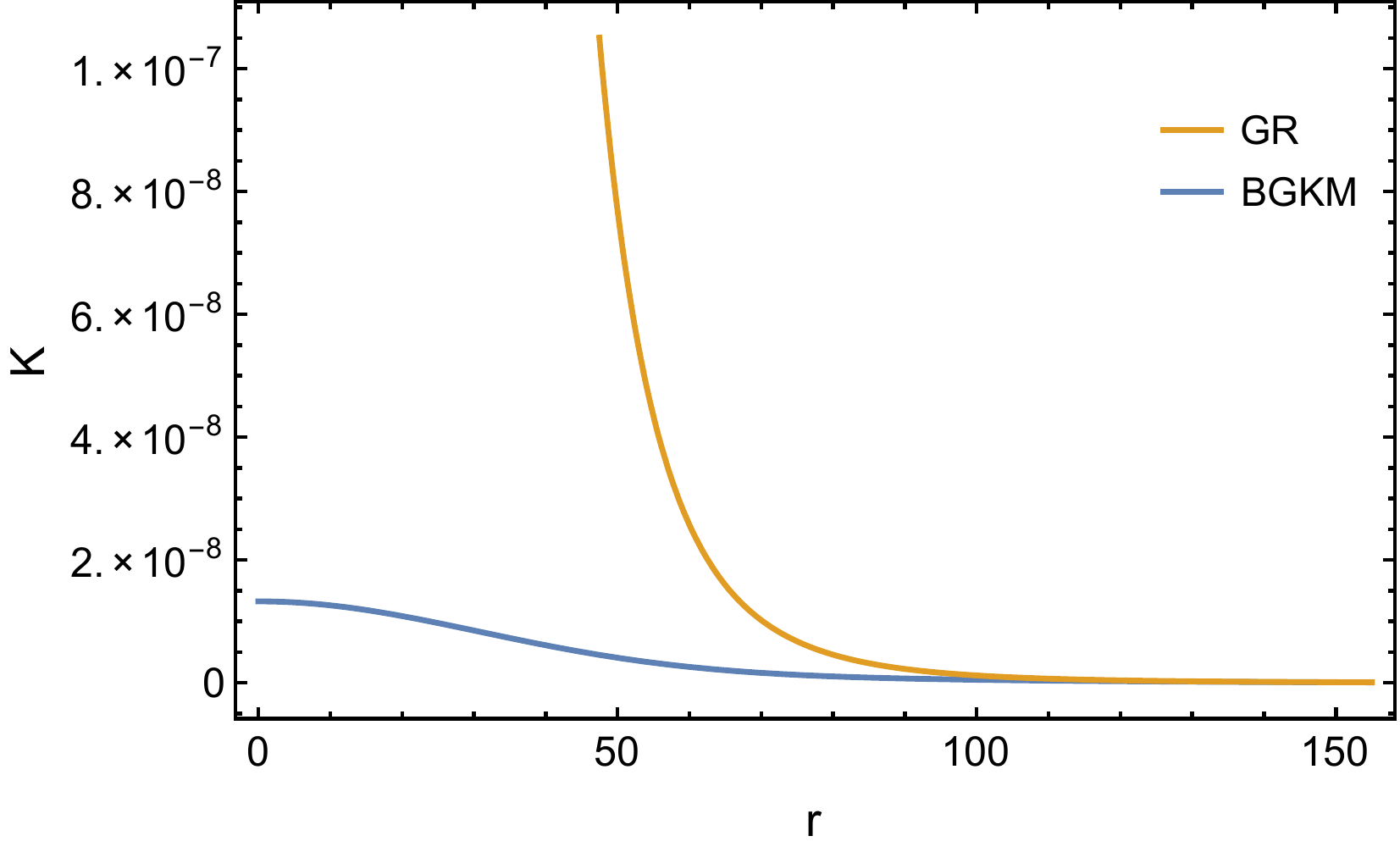}
		\label{fig.2.c}}
	\protect\caption{We have shown the behavior of some tensors in both the cases of BGKM (blue line) and GR (orange line): (a) $\mathcal{R}_{0101}$ component of the Riemann tensor; (b) $\mathcal{C}_{0101}$ component of the Weyl tensor; and (c) the Kretschamnn invariant tensor $\mathcal{K}$. We have set $M_p=1,~m=5$ and $M_s=10^{-3/2}$. In these units the Schwarzschild radius is $r_{sch}=10$. Note that, unlike in GR where the linearized approximation would break down for $r<10$, in the case of BGKM we can smoothly approach $r=0$, provided $mM_s<M_p^2.$ From these three plots it is obvious that the presence of non-locality helps to avoid the curvature-singularities at the origin in the linearized regime.}\label{fig2}
\end{figure}
From Eqs. \eqref{eq:17}-\eqref{eq:20} we can already understand an important feature of the BGKM gravity: unlike Einstein's GR, where Schwarzschild metric yields a vacuum solution for which the Ricci tensor and the Ricci scalar vanish, the corresponding metric in Eq. \eqref{eq:5} is not a vacuum solution for the BGKM gravity, at the linearized level. Physically what happens is the following: far from the source, $ r > 2/M_s$, gravity is very well described by the predictions of Einstein's GR, and in the IR limit all the infinite derivate contributions becomes negligible for $r > 2/M_s$, but as soon as we move into the non-local region, i.e. $r<2/M_s$, the smearing effect of the source induced by the non-locality becomes relevant and gives a non-zero value to the components of the Ricci tensor and the Ricci scalar, which exhibit a Gaussian behavior. 

Such an interesting feature also manifests that in the BGKM gravity the Weyl tensor does not coincide with the Riemann tensor, which, unlike in Einstein's GR, is not traceless. Indeed, the non-vanishing components of the Weyl tensor are given by:
\begin{equation}
\begin{array}{rl}
\mathcal{C}_{0101}= & \displaystyle \frac{e^{-\frac{M_s^2r^2}{4}}Gm}{3\sqrt{\pi}r^3}\left[6M_sr+M_s^3r^3\right.\\
& \displaystyle \,\,\,\,\,\,\,\,\,-\left.6\sqrt{\pi}e^{\frac{M_s^2r^2}{4}}{\rm Erf}\left(\frac{M_sr}{2}\right)\right],\\
\mathcal{C}_{0303}= & \displaystyle -\mathcal{C}_{1313}=\mathcal{C}_{0202}{\rm sin}^2\theta=-\mathcal{C}_{1212}{\rm sin}^2\theta\\
= & \displaystyle -\frac{e^{-\frac{M_s^2r^2}{4}}Gm{\rm sin}^2\theta}{6\sqrt{\pi}r}\left[6M_sr+M_s^3r^3\right.\\
& \displaystyle \,\,\,\,\,\,\,\,\,-\left.6\sqrt{\pi}e^{\frac{M_s^2r^2}{4}}{\rm Erf}\left(\frac{M_sr}{2}\right)\right],\\
\mathcal{C}_{2323}= & \displaystyle -\frac{e^{-\frac{M_s^2r^2}{4}}Gmr{\rm sin}^2\theta}{3\sqrt{\pi}}\left[6M_sr+M_s^3r^3\right.\\
& \displaystyle \,\,\,\,\,\,\,\,\,-\left.6\sqrt{\pi}e^{\frac{M_s^2r^2}{4}}{\rm Erf}\left(\frac{M_sr}{2}\right)\right].
\label{eq:21}
\end{array}
\end{equation}
First of all, note that the result in Eq. \eqref{eq:21} is consistent with that of Einstein's GR, indeed in the IR limit, i.e. $M_sr > 2$, we recover the components in Eq. \eqref{eq:12}. A very important feature now emerges - in the non-local region $M_sr<1$, all the components of the Weyl tensor go quadratically to zero as $r\rightarrow 0$:
\begin{equation}
\mathcal{C}_{\mu\nu\rho\sigma}\sim 0,
\label{eq:22}
\end{equation}
as evident from Fig. 2.b. for the component $\mathcal{C}_{0101}$.

The result in Eq. \eqref{eq:22} implies that the BGKM metric in Eqs.~(\ref{eq:5},\ref{eq:6}) becomes {\it conformally flat} in the non-local region $r<2/M_s.$ In fact, the metric in Eq.\eqref{eq:5} in the non-local region assumes the following form:
\begin{equation}
ds^{2}=-(1-2A)dt^{2}+(1+2A)\left[dr^2+r^2\Omega^2\right] \label{eq:23},
\end{equation}
where $A$ has been defined in Eq. \eqref{const.pot.}. In this region, $r< 2/M_s$, we can easily perform a coordinate transformation from isotropic coordinates $(t,r,\theta,\varphi)$ to Schwarzschild coordinates $(t,R,\theta,\varphi)$, where $R$ is the polar radial coordinate, indeed, the following relation holds~\footnote{Note that the metric in Eq. \eqref{eq:5} cannot be rewritten in Schwarzschild coordinates due to the fact that the coordinate transformation $R^2=(1-2\phi(r))r^2$ is not analytically invertible, but we can easily invert it in the regime of non-locality.}:
\begin{equation}
R=\sqrt{1+2A}r \label{eq:24},
\end{equation}
and we obtain the same metric in the Schwarzschild coordinates:
\begin{equation}
ds^{2}=-(1-2A)dt^{2}+dR^{2}+R^2d\Omega^2 \label{eq:25},
\end{equation}
which is clearly not Minkowski due to the presence of the factor $1-2A$ in front of the time component. However, one can easily show that the metric in Eq. \eqref{eq:23} is conformally flat by introducing a new time coordinate that we call {\it conformal time}, where $A<1$, since we are working in the weak-field regime:
\begin{equation}
\tau=\sqrt{\frac{1-2A}{1+2A}}t \label{eq:26},
\end{equation}
so that the metric in Eq. \eqref{eq:23} in the non-local regime, $ r< 2/M_s$, can be recast as:
\begin{equation}
\begin{array}{rl}
ds^{2}= & (1+2A)\left(-d\tau^{2}+dr^{2}+r^2d\Omega^2\right)\\ \label{eq:27}
= & F\eta,
\end{array}
\end{equation}
where $\eta$ is the Minkowski metric, and $F:=1+2A>0$ is the conformal factor corresponding to the conformal transformation in Eq. \eqref{eq:26}.

Now, it only remains to compute the squares of the curvatures. They are given by:
\begin{equation}
\begin{array}{rl}
\mathcal{R}^2= & \mathcal{R}_{\mu\nu}\mathcal{R}^{\mu\nu}= \displaystyle \frac{e^{-\frac{M_s^2r^2}{2}}G^2m^2M_s^6}{\pi},\\
\mathcal{C}^2= & \displaystyle -\frac{4}{3}\frac{e^{-\frac{M_s^2r^2}{2}}G^2m^2}{\pi r^6}\left[6M_sr+M_s^3r^3\right.\\
& \displaystyle \,\,\,\,\,\,\,\,\,-\left.6\sqrt{\pi}e^{\frac{M_s^2r^2}{4}}{\rm Erf}\left(\frac{M_sr}{2}\right)\right]^2,\\
\mathcal{K}= &\displaystyle -\frac{e^{-\frac{M_s^2r^2}{2}}G^2m^2}{3\pi r^6}\left\lbrace 5M_s^6r^6+ 4\left[6M_sr+M_s^3r^3\right.\right.\\
& \displaystyle \,\,\,\,\,\,\,\,\,-\left.\left.6\sqrt{\pi}e^{\frac{M_s^2r^2}{4}}{\rm Erf}\left(\frac{M_sr}{2}\right)\right]^2\right\rbrace.
\label{eq:28}
\end{array}
\end{equation}
First of all, the Ricci squared and the Ricci scalar squared tend consistently to zero in the IR limit, $r> 2/M_s$, while the invariants $\mathcal{K}$ and $\mathcal{C}^2$ recover the ones in Eq. \eqref{eq:13}. Moreover, non-locality  in the BGKM gravity is such that we can avoid the singularities of the invariant tensors at the origin, indeed in the regime of non-locality, $M_sr<2,$ $\mathcal{C}^2$ goes to zero, while the others tend to finite constant values:
\begin{equation}
\begin{array}{ll}
\displaystyle \mathcal{R}^2\sim\frac{G^2m^2M_s^6}{\pi}, &\displaystyle\,\,\,\,\,\,\,\,\,\mathcal{R}_{\mu\nu}R^{\mu\nu}\sim\frac{G^2m^2M_s^6}{\pi},\\
\displaystyle\mathcal{K}\sim\frac{5G^2m^2M_s^6}{3\pi}, &\displaystyle\,\,\,\,\,\,\,\,\,\mathcal{C}^2\sim0.
\end{array}
\label{eq:29}
\end{equation}
In Fig. 2.c. we have shown the behavior of the Kretschmann invariant for both the cases of BGKM and GR.

We have learnt that the non-local region of size $r \sim 2/M_s$ can be approximatively described by a conformally flat manifold with non-negative constant curvatures.
For completeness, it can be easily checked that the invariant tensors in Eq. \eqref{eq:28} consistently satisfy the following well-known relation:
\begin{equation}
\mathcal{C}^2= \frac{\mathcal{R}^2}{3}-2\mathcal{R}_{\mu\nu}\mathcal{R}^{\mu\nu}+\mathcal{K}. \label{eq:30}
\end{equation}

\section{Towards static non-linear solution of BGKM gravity in the UV regime}

The results we have obtained so far hold true in the linearized regime, where we have considered the first order perturbations around Minkowski spacetime. Despite the fact that we are neglecting non-linear contributions, the analysis made in the previous sections allow us to understand what would happen if we were to solve the full non-linear field equations corresponding to the action in Eq. \eqref{eq:1}, derived in Ref.~\cite{Biswas:2013cha}, see the full non-linear field equations (Eqs. 52-58 of Ref.~\cite{Biswas:2013cha}). 

The field equations of BGKM gravity are extremely cumbersome, and here we express only the trace equation for our discussion, and also for the brevity:
\[
\label{trace}
\begin{aligned}
P&=-R+12\square{\cal F}_{1}(\Box_s)R+2\square({\cal
F}_{2}(\Box_s)R)\\&
+4\nabla_{\mu}\nabla_{\nu}({\cal
F}_{2}(\Box_s)R^{\mu\nu})
+2(\Omega_{1\sigma}^{\;\sigma}+2\bar{\Omega}_{1})\\&
+2(\Omega_{2\sigma}^{\;\sigma}
+2\bar{\Omega}_{2})+2(\Omega_{3\sigma}^{\;\sigma}+2\bar{\Omega}_{3}
)-4\Delta_{2\sigma}^{\;\sigma}\\&
-8\Delta_{3\sigma}^{\;\sigma}=T\equiv g_{\rho\sigma}T^{\rho\sigma}\ ,
\end{aligned}
\]
where $\Omega_{1\sigma}^{\;\sigma},~\bar{\Omega}_{1}$ are functions of ${\cal R}$ and covariant derivatives, ${\cal R}^{(m)}\equiv \Box^{m}_s {\cal R}$,
$\Omega_{2\sigma}^{\;\sigma},~\bar{\Omega}_{2},~\Delta_{2\sigma}^{\;\sigma}$ are functions of ${\cal R}^{\mu\nu}$ and covariant derivatives, 
${\cal R}^{\mu\nu (m)}\equiv \Box^{m}_s {\cal R}^{\mu\nu}$,
$\Omega_{3\sigma}^{\;\sigma},~\bar{\Omega}_{3},~\Delta_{3\sigma}^{\;\sigma}$ are similarly functions of ${\cal C}^{\mu\nu\lambda\sigma}$ and covariant derivatives, ${\cal C}^{\mu\nu \lambda\sigma(m)}\equiv \Box^{m}_s {\cal C}^{\mu\nu\lambda\sigma}$. The exact expressions are derived in Ref.~\cite{Biswas:2013cha}. 

The UV regime of the equations of motion are determined by the non-local terms, i.e. all the terms except the first term, $R$, which is the local contribution and determines the recovery of the IR results, i.e. when
$r> 2/M_s$. In the latter case, in the local regime the solution will exactly yield the Schwarzschild geometry. For $r <2/M_s$, rest of the terms dominate the equations of motion.

Note that any full non-linear solution of the above Eq.~(\ref{trace}) must have a smooth limit such that our results would be recovered in the linearized regime (even in the UV), which are:
\begin{itemize}
\item The Ricci tensor and the Ricci scalar are not vanishing, and the Riemann tensor is not traceless, inside $r < 2/M_s$;
\item The Weyl tensor vanishes as $r <2/M_s$.
\end{itemize}
These conclusions are indeed in stark contrast with respect to Einstein's GR, where the vacuum solution of the field equations, i.e. ${\cal R}=0,~{\cal R}^{\mu\nu}=0$ give rise to $1/r$ metric potential in both linear and non-linear regimes.

In the UV regime, for $r< 1/M_s$, we may proceed by {\it reductio ad absurdum}: 
\begin{itemize}

\item if the full non-linear solution satisfies $\mathcal{R_{\mu\nu}=\mathcal{R}}=0,$ then the metric potential should yield 
$1/r$-fall of the metric potential in the linear, and in the non-linear regime, even in the UV region, irrespective of the action.

\item Furthermore, looking at Eq. (\ref{trace}), the only non-vanishing contribution must then arise from the Weyl part, involving  $\Omega_{3\sigma}^{\;\sigma},~\bar{\Omega}_{3},~\Delta_{3\sigma}^{\;\sigma}$. Indeed, there are two possibilities, which would yield the Weyl contribution to zero for the vacuum solution; either ${\cal C}^{\mu\nu\rho\sigma}  \rightarrow 0$ for $r< 2/M_s$, or a fine cancellation between terms
$\Omega_{3\sigma}^{\;\sigma},~\bar{\Omega}_{3},~\Delta_{3\sigma}^{\;\sigma}$, while the latter is indeed highly unlikely and fine-tuned possibility.

\end{itemize}

Therefore, we may conclude that  in the full non-linear scenario, the metric solution has to be such that the Ricci tensor and the Ricci scalar are non-zero, and the Riemann is not traceless, so that it does not coincide with the Weyl, and $1/r$ fall of the gravitational potential cannot be a metric solution for the full non-linear theory at least in the UV regime, where $r < 2/M_s$.  

Having argued so, there remains a possibility that ${\cal C}^{\mu\nu\rho\sigma}\neq 0$, as we approach $r\rightarrow 0$ inside $r< 2/M_s$, but this would require a very unlikely cancellation between the terms involving $\Omega_{3\sigma}^{\;\sigma},~\bar{\Omega}_{3},~\Delta_{3\sigma}^{\;\sigma}$.

In this respect, let us point towards an interesting possibility suggested in Ref.~\cite{Koshelev:2017bxd}, where it has been pointed out that the astrophysical solar massive blackholes within BGKM theory can have no singularity and event horizon, provided the scale of non-locality itself shifts from $M_s \rightarrow M_s/\sqrt{N}$, where $N$ being the number of {\it plaquettes}, or the degrees of freedom associated with the baryons and gravitons at the last stages of the formation of a blackhole. For sufficiently large $N$, the scale of non-locality $M_s$ (which signifies the scale of quantum gravity) shifts towards the IR scale, where in the region inside $r <r_\ast\sim r_{sch}$, the potential remains linear forever given by Eq.~(\ref{eq:6}), and satisfies all our equations of section IV.

\section{Conclusions}

We briefly summarize the main results obtained here. We have computed all the linearized curvature tensors for the metric in Eq. \eqref{eq:5} for the BGKM action, which highlights at a classical level crucial properties of the non-local nature of the gravitational interaction. We have found that the Ricci tensor and the Ricci scalar are not vanishing in the limit $r< 2/M_s$, as a consequence of the non-local smearing of the source induced by the infinite derivatives. Such a feature implies that the metric is not a vacuum solution and the trace of the Riemann tensor is also non-vanishing unlike in the case of Schwarzschild metric in Einstein's GR. Furthermore, the components of the Weyl tensor {\it do not coincide} with the components of the Riemann tensor. 

Moreover, we have shown that in the region of non-locality, i.e. $r< 2/M_s$, Riemann, Ricci and scalar tensors tend to finite constant values, while the Weyl tensor tends to zero quadratically in $r$, implying that in such a non-local region, of size $\sim 2/M_s$, the spacetime metric becomes conformally flat with a non-negative constant curvature. 
In the end, we have also computed all the possible invariants, and showed explicitly that the Kretschmann invariant is non-singular at $r=0$, as a consequence of the non-local interaction of short-distance gravity in BGKM. All these invariants have been computed in the linear regime, see Eqs. (\ref{eq:5},\ref{eq:6}).

We have briefly alluded towards seeking the non-linear solution for the BGKM action. Indeed, now there are hints that even in the full non-linear theory we should not expect $1/r$-metric potential in the UV regime, for $r< 2/M_s$, due to the fact that any non-linear solution must interpolate smoothly to the linear solution.
Of course, such arguments should be manifestly tested at the level of the field equations for the full action given in Ref.~\cite{Biswas:2013cha}. 

This work provides a solid foundation for addressing the full non-linear solution in the near future. Furthermore, all our results are valid for a static metric, it is indeed possible to perform similar computations for the rotating metric in the BGKM gravity, which has been recently discussed in Ref.~\cite{Cornell:2017irh}, at the linearized level, which has no horizon and no ring-like singularity.

\acknowledgements
AK is supported by FCT Portugal investigator project IF/01607/2015, FCT Portugal fellowship SFRH/BPD/105212/2014, in  by FCT Portugal grant UID/MAT/00212/2013, and COST Action CA15117 (CANTATA).


\begin{thebibliography}{1}
	
	\bibitem{-C.-M.}C. M. Will, Living Rev. Rel. 17, 4
	(2014) {[}arXiv:1403.7377 {[}gr-qc{]}{]}.
	
	\bibitem{-B.-P.}B. P. Abbott et al. {[}LIGO Scientific
	and Virgo Collaborations{]}, Phys. Rev. Lett. 116 (2016) no.6, 061102.
	
	\bibitem{-D.-J.}D. J. Kapner, T. S. Cook, E. G. Adelberger,
	J. H. Gundlach, B. R. Heckel, C. D. Hoyle and H. E. Swanson, Phys.
	Rev. Lett. \textbf{98} (2007) 021101.
	
	\bibitem{-K.-S.}K. S. Stelle, Phys. Rev. D \textbf{16} (1977)
	953.
	
	
\bibitem{Biswas:2005qr} 
  T.~Biswas, A.~Mazumdar and W.~Siegel,
  ``Bouncing universes in string-inspired gravity,''
  JCAP {\bf 0603}, 009 (2006)
  [hep-th/0508194].
	
	
\bibitem{Biswas:2011ar} 
T.~Biswas, E.~Gerwick, T.~Koivisto and A.~Mazumdar,
``Towards singularity and ghost free theories of gravity,''
Phys.\ Rev.\ Lett.\  {\bf 108}, 031101 (2012).
	
\bibitem{Biswas:2016etb} 
T.~Biswas, A.~S.~Koshelev and A.~Mazumdar,
``Gravitational theories with stable (anti-)de Sitter backgrounds,''
Fundam.\ Theor.\ Phys.\  {\bf 183}, 97 (2016).
T.~Biswas, A.~S.~Koshelev and A.~Mazumdar,
``Consistent higher derivative gravitational theories with stable de Sitter and anti?de Sitter backgrounds,''
Phys.\ Rev.\ D {\bf 95}, no. 4, 043533 (2017).
	
\bibitem{-Yu.-V.}Yu. V. Kuzmin, Yad. Fiz. 50, 1630-1635 (1989).

\bibitem{Tomboulis}
E. Tomboulis, Phys. Lett. B 97, 77 (1980). E. T. Tomboulis, Renormalization And Asymptotic Freedom In Quantum Gravity, In *Christensen, S.m. ( Ed.): 
Quantum Theory Of Gravity*, 251-266. E. T. Tomboulis, Superrenormalizable gauge and gravitational theories, hep- th/9702146;
	
\bibitem{Tseytlin:1995uq} 
  A.~A.~Tseytlin,
  ``On singularities of spherically symmetric backgrounds in string theory,''
  Phys.\ Lett.\ B {\bf 363}, 223 (1995)
  [hep-th/9509050].

\bibitem{Siegel:2003vt}
  W.~Siegel,
  ``Stringy gravity at short distances,''
  hep-th/0309093.

\bibitem{polchinski}
J. Polchinski. String theory. Vol. 2: Superstring theory and beyond - 1998. Univ. Pr.. Cambridge, UK: r. 531p.

\bibitem{Witten:1985cc}
  E.~Witten,
  ``Noncommutative Geometry and String Field Theory,''
  Nucl.\ Phys.\ B {\bf 268} (1986) 253.
  P.~G.~O.~Freund and M.~Olson,
  Phys.\ Lett.\ B {\bf 199}, 186 (1987).
  P.~H.~Frampton and Y.~Okada,
  ``Effective Scalar Field Theory of $P^-$adic String,''
  Phys.\ Rev.\ D {\bf 37}, 3077 (1988).
  
  \bibitem{Siegel:1988yz} 
  W.~Siegel,
  ``Introduction to string field theory,''
  Adv.\ Ser.\ Math.\ Phys.\  {\bf 8}, 1 (1988)
  [hep-th/0107094].
		
	
\bibitem{Biswas:2013kla} 
T.~Biswas, T.~Koivisto and A.~Mazumdar,
``Nonlocal theories of gravity: the flat space propagator,''
arXiv:1302.0532 [gr-qc].

\bibitem{Conroy:2015nva} 
  A.~Conroy, A.~Mazumdar, S.~Talaganis and A.~Teimouri,
  ``Nonlocal gravity in D dimensions: Propagators, entropy, and a bouncing cosmology,''
  Phys.\ Rev.\ D {\bf 92}, no. 12, 124051 (2015)
  [arXiv:1509.01247 [hep-th]].
	  
\bibitem{Buoninfante} L. Buoninfante, Master's Thesis (2016),
{[}arXiv:1610.08744v4 {[}gr-qc{]}{]}.
	
\bibitem{Tomboulis:2015}	
E.~T.~Tomboulis,
	``Nonlocal and quasilocal field theories,''
	Phys.\ Rev.\ D {\bf 92}, no. 12, 125037 (2015).
	
\bibitem{Modesto}L. Modesto, Phys. Rev. D 86, 044005 (2012), {[}arXiv:1107.2403
	{[}hep-th{]}{]}.
	
\bibitem{Edholm:2016hbt} 
	J.~Edholm, A.~S.~Koshelev and A.~Mazumdar,
	``Behavior of the Newtonian potential for ghost-free gravity and singularity-free gravity,''
	Phys.\ Rev.\ D {\bf 94}, no. 10, 104033 (2016).
	V. P. Frolov and A. Zelnikov,  Phys. Rev. D 93, no. 6, 064048 (2016).
	
\bibitem{Talaganis:2014ida} 
	S.~Talaganis, T.~Biswas and A.~Mazumdar,
	``Towards understanding the ultraviolet behavior of quantum loops in infinite-derivative theories of gravity,''
	Class.\ Quant.\ Grav.\  {\bf 32}, no. 21, 215017 (2015).
S.~Talaganis and A.~Mazumdar,
  ``High-Energy Scatterings in Infinite-Derivative Field Theory and Ghost-Free Gravity,''
  Class.\ Quant.\ Grav.\  {\bf 33}, no. 14, 145005 (2016)
  doi:10.1088/0264-9381/33/14/145005
  [arXiv:1603.03440 [hep-th]].

\bibitem{Frolov}V. P. Frolov,  Phys. Rev. Lett. 115, no. 5, 051102 (2015).

\bibitem{Buoninfante:2017kgj} 
  L.~Buoninfante, G.~Lambiase and A.~Mazumdar,
  ``Quantum solitonic wave-packet of a meso-scopic system in singularity free gravity,''
  arXiv:1708.06731 [quant-ph].
   L.~Buoninfante, G.~Lambiase and A.~Mazumdar,
  ``Quantum spreading of a self-gravitating wave-packet in singularity free gravity,''
  Eur.\ Phys.\ J.\ C {\bf 78}, no. 1, 73 (2018)
  [arXiv:1709.09263 [gr-qc]].
  
  \bibitem{Bose:2017nin} 
  S.~Bose {\it et al.},
  ``Spin Entanglement Witness for Quantum Gravity,''
  Phys.\ Rev.\ Lett.\  {\bf 119}, no. 24, 240401 (2017)
  doi:10.1103/PhysRevLett.119.240401
  [arXiv:1707.06050 [quant-ph]].


\bibitem{Koivisto}
T.~Biswas, T.~Koivisto and A.~Mazumdar,
	``Towards a resolution of the cosmological singularity in non-local higher derivative theories of gravity,''
	JCAP {\bf 1011}, 008 (2010).


\bibitem{Koshelev:2012qn} 
	A.~S.~Koshelev and S.~Y.~Vernov,
	``On bouncing solutions in non-local gravity,''
	Phys.\ Part.\ Nucl.\  {\bf 43}, 666 (2012).
	T.~Biswas, A.~S.~Koshelev, A.~Mazumdar and S.~Y.~Vernov,
	``Stable bounce and inflation in non-local higher derivative cosmology,''
	JCAP {\bf 1208}, 024 (2012).
	
\bibitem{Conroy:2014dja} 
  A.~Conroy, A.~S.~Koshelev and A.~Mazumdar,
  ``Geodesic completeness and homogeneity condition for cosmic inflation,''
  Phys.\ Rev.\ D {\bf 90}, no. 12, 123525 (2014)
  [arXiv:1408.6205 [gr-qc]].
A.~Conroy, A.~S.~Koshelev and A.~Mazumdar,
  ``Defocusing of Null Rays in Infinite Derivative Gravity,''
  JCAP {\bf 1701}, no. 01, 017 (2017)
  [arXiv:1605.02080 [gr-qc]].

\bibitem{Penrose:1964wq} 
  R.~Penrose,
  ``Gravitational collapse and space-time singularities,''
  Phys.\ Rev.\ Lett.\  {\bf 14}, 57 (1965).
  
  \bibitem{Hawking:1973uf} 
  S.~W.~Hawking and G.~F.~R.~Ellis,
  ``The Large Scale Structure of Space-Time,''
	
\bibitem{Raychaudhuri:1953yv} 
  A.~Raychaudhuri,
  Phys.\ Rev.\  {\bf 98}, 1123 (1955).
  	

\bibitem{Frolov:2015bia} 
  V.~P.~Frolov, A.~Zelnikov and T.~de Paula Netto,
  JHEP {\bf 1506}, 107 (2015)
  [arXiv:1504.00412 [hep-th]].
  
  \bibitem{Frolov:2015usa} 
  V.~P.~Frolov and A.~Zelnikov,
  Phys.\ Rev.\ D {\bf 93}, no. 6, 064048 (2016)
  [arXiv:1509.03336 [hep-th]].


\bibitem{Biswas:2013cha} 
T.~Biswas, A.~Conroy, A.~S.~Koshelev and A.~Mazumdar,
``Generalized ghost-free quadratic curvature gravity,''
Class.\ Quant.\ Grav.\  {\bf 31}, 015022 (2014),
Erratum: [Class.\ Quant.\ Grav.\  {\bf 31}, 159501 (2014)].
 [arXiv:1308.2319 [hep-th]].
	

\bibitem{Koshelev:2017bxd} 
  A.~S.~Koshelev and A.~Mazumdar,
  ``Do massive compact objects without event horizon exist in infinite derivative gravity?,''
  Phys.\ Rev.\ D {\bf 96}, no. 8, 084069 (2017)
  [arXiv:1707.00273 [gr-qc]].



\bibitem{Cornell:2017irh} 
  A.~S.~Cornell, G.~Harmsen, G.~Lambiase and A.~Mazumdar,
  ``Rotating metric in Non-Singular Infinite Derivative Theories of Gravity,''
  arXiv:1710.02162 [gr-qc].










		
		
		
	
		
	
	
	
	
	
		
		
		

 	

	
	
	
\end{thebibliography}
\end{document}